\magnification=\magstep1

\def\chaphead{}
\def\ni{\noindent}

\font\tfont=cmbxti10
\font\eightrm=cmr8
\font\eightit=cmti8
\font\sixrm=cmr6
\font\eightmit=cmmi8
\font\sixmit=cmmi6
\def\absmath{\textfont0=\eightrm \scriptfont0=\sixrm
	      \textfont1=\eightmit \scriptfont1=\sixmit}
\def\absfont{\let\rm=\eightrm \let\it=\eightit \rm\absmath}
\font\twelverm=cmr12
\font\twelveit=cmti12
\font\tenrm=cmr10
\font\twelvemit=cmmi12
\font\tenmit=cmmi10
\def\regmath{\textfont0=\twelverm \scriptfont0=\tenrm
	      \textfont1=\twelvemit \scriptfont1=\tenmit}
\def\peterfont{\let\rm=\twelverm \let\it=\twelveit \rm\regmath}
%
%

\newfam\vecfam

\textfont\vecfam=\tfont \scriptfont\vecfam=\seveni
\scriptscriptfont\vecfam=\fivei


\def\spose#1{\hbox to 0pt{#1\hss}}

\font\eightrm=cmr8

\def\s{\ifmmode \widetilde \else \~\fi} 
     
\def\section{\S}
\newcount\notenumber
\notenumber=1
\newcount\eqnumber
\eqnumber=1
\newcount\fignumber
\fignumber=1
\newbox\abstr


\def\s{{\rm\,s}}

\def\note#1{\footnote{$^{\the\notenumber}$}{#1}\global\advance\notenumber by 1}
\def\foot#1{\raise3pt\hbox{\eightrm \the\notenumber}
     \hfil\par\vskip3pt\hrule\vskip6pt
     \noindent\raise3pt\hbox{\eightrm \the\notenumber}
     #1\par\vskip6pt\hrule\vskip3pt\noindent\global\advance\notenumber by 1}

\def\abstract#1{\setbox\abstr=\vbox{\hsize 5.0truein{\par\noindent#1}}
    \centerline{ABSTRACT} \vskip12pt \hbox to \hsize{\hfill\box\abstr\hfill}}
     
\def\Dt{\spose{\raise 1.5ex\hbox{\hskip3pt$\mathchar"201$}}}    
\def\dt{\spose{\raise 1.0ex\hbox{\hskip2pt$\mathchar"201$}}}    

\def\new{{\rm\chaphead\the\eqnumber}\global\advance\eqnumber by 1}
\def\ref#1{\advance\eqnumber by -#1 \chaphead\the\eqnumber
     \advance\eqnumber by #1 }
\def\last{\advance\eqnumber by -1 {\rm\chaphead\the\eqnumber}\advance
     \eqnumber by 1}
\def\eqnam#1{\xdef#1{\chaphead\the\eqnumber}}
     
\def\nfig{\chaphead\the\fignumber\global\advance\fignumber by 1}
\def\nfiga#1{\chaphead\the\fignumber{#1}\global\advance\fignumber by 1}
\def\rfig#1{\advance\fignumber by -#1 \chaphead\the\fignumber
     \advance\fignumber by #1}
\def\fignam#1{\xdef#1{\chaphead\the\fignumber}}

\def\lta{\mathrel{\spose{\lower 3pt\hbox{$\mathchar"218$}}
     \raise 2.0pt\hbox{$\mathchar"13C$}}}
\def\gta{\mathrel{\spose{\lower 3pt\hbox{$\mathchar"218$}}
     \raise 2.0pt\hbox{$\mathchar"13E$}}}
     

\magnification=\magstep1
\parskip=3pt


\font\syvec=cmbsy10                        

\def\bnabla{\hbox{{\syvec\char114}}}       

     


\def\ni{\noindent}
\def\refind{\noindent \hangindent=2pc \hangafter=1}
\overfullrule=0pt

\centerline{\bf Gas accretion in a clumpy disk with application to AGNs}
\bigskip
\centerline{\bf Pawan Kumar$^\dagger$}
\medskip
\centerline{Institute for Advanced Study, Princeton, NJ 08540}
\bigskip\bigskip
\centerline{\bf Abstract} 
\bigskip

We analyze the collective gravitational interaction among gas clouds in 
the inner regions of galactic disks and find that it leads to accretion
at a rate $\sim M_{mc}\Omega (M_{mc}/M_t)^2$; where $M_{mc}$ is the molecular 
mass of the disk, $M_t$ is sum of the central plus any axisymmetrically 
distributed mass, and $\Omega$ is the mean angular speed of clumps. 
We discuss applications of this result to the mega-maser galaxy NGC 4258,
for which we have observational evidence that the maser spots are
concentrated in a thin molecular disk which is clumpy, and find the 
accretion rate to be $\sim 1.5\times10^{-3}$ M$_\odot$ yr$^{-1}$.
If the gravitational energy release of this inward falling gas were to be 
radiated away efficiently, then the resulting luminosity would greatly
exceed the observed central luminosity of NGC 4258, indicating that
most of the thermal energy of the gas is advected with the flow into 
the blackhole as proposed by Lasota et al. (1996).

The gravitational interactions among molecular clouds lying within the
inner kpc of our galaxy give an accretion rate of $\sim 10^{-5}$ 
M$_\odot$ yr$^{-1}$, which is consistent with the value obtained by
Narayan et al. (1995) by fitting the spectrum of Sagitarrius A$^*$. 
We also discuss possible application of this work to quasar evolution.

\medskip
\ni{\it Subject heading:} accretion disks -- galaxies: individual (NGC 4258)

\vskip 9.3truecm
\noindent $^\dagger$ Alfred P. Sloan Fellow

\vfill\eject

\centerline{\bf 1. Introduction}
\medskip The gas in the inner parsec of a number of active 
galaxies appears to be clumpy and concentrated in a thin disk. The 
best evidence for this comes from the observation of discrete spots of 
water maser emission in active galaxies such as NGC 4258, 1068, 4945 
(cf. Watson \& Wallin 1994, Greenhill et al. 1995, Miyoshi et al. 1995, 
Greenhill et al. 1996 \& 1997, Braatz et al. 1996, and references
therein). The luminosity of the central source in these systems is 
believed to be due to the release of gravitational energy of 
infalling gas. The gas accretion could be a result of turbulent 
viscosity in the disk, however its effectiveness in a cold thin 
disk consisting of discrete molecular clouds in very uncertain. 

Here we consider another, more likely, process 
driving gas accretion -- the gravitational interaction among clumps, 
or clouds, excites epicyclic oscillation, causing clouds to collide 
(inelastically) and as a result some fraction of them fall inward.
We treat clumps as point mass when considering their gravitational
interactions, which excite epicyclic oscillations. The finite clump
size is taken into consideration when determining the frequency of
physical collisions among clumps. We assume that there is no dipole or
low multipole order coherent density structure in the disk which 
induces efficient inward flow of gas. The effect of spiral density
wave on accretion in NGC 4258 has been considered by Maoz and McKee 
in a recent paper. The process we consider has some similarity to the
work of Goldreich and Tremaine (1978), however unlike the case of the
planetary ring systems considered by these authors the gravitational
interactions between clouds are very important and is considered 
in some detail here.

There is another motivation for considering the gravitational interaction
among gas clumps. The mechanism for the inward transport of gas within 
the inner kpc of active galaxies (AGNs) to supply fuel for the central 
activity, is poorly understood. Tidal torques from neighboring galaxies, 
even in close encounters, are believed to be ineffective at this 
distance (but it can be important in transporting gas from several 
kilo-parsecs to within about a kpc). Moreover, recent 
observations of a sample of spiral galaxies show that the central 
activity is uncorrelated with the strength of the galactic bar
(e.g. Ho, Filippenko and Sargent, 1997), which suggests that bars are 
not effective in inducing radial inflow of gas either. We estimate the
efficiency of gas transport, within the inner kpc of galactic disks, due
to gravitational interaction between clumps in the disk (Shlosman \& 
Begelman 1987, also investigated this mechanism; see also Paczynski 1978, 
and Shlosman, Begelman \& Frank 1990 and references therein).

We provide an estimate of gas accretion rate in a clumpy disk in \S 2 and 
apply it to the water maser galaxies and fueling of AGNs in \S 3.

\bigskip
\centerline{\bf 2. Accretion in a clumpy disk}

The accretion rate due to gravitational interaction among clumps is 
derived in \S2.2.1. Although pressure forces are ignored in this paper, 
the method of \S2.1 can be generalized to include pressure of the inter-clump
medium. A physical, but crude, derivation of the main result is also presented
in \S2.2.

\medskip
\centerline{\S2.1 Basic equations}
\smallskip
\medskip We consider a thin molecular disk which contains clumps of gas.  
We assume that the disk does not have some unstable dipole or other low 
order long wavelength mode which is excited to large amplitude causing 
efficient transport of angular momentum outward and enabling gas to 
fall inward.  However, we assume that some local instability,
gravitational or thermal, exists and is responsible for the lumpiness 
of the disk. The clumps move on nearly circular orbits under the 
influence of an axisymmetric potential $\psi(r)$ and the gravitational 
attraction of other clumps. The equation of motion of the clumps, neglecting 
pressure forces, is given by

$$ {d^2 {\bf r} \over dt^2} = -\bnabla\Psi - G\, \bnabla \int d^2 r_1\; 
   {\sigma(t, {\bf r_1})\over |{\bf r_1 - r}|}, \eqno(\new)$$
where $\sigma(t,{\bf r})$ is the surface mass density of gas, and 
${\bf r}$ is the center of mass of the clump. Expanding the angular 
dependence of the surface density $\sigma(t,r,\phi)$ in a Fourier 
series, and substituting the spherical harmonic expansion for $|{\bf 
r_1 - r}|^{-1}$ in the equation above, we obtain

$$ {d^2 {\bf r} \over dt^2} = -\bnabla\Psi - 2\pi G \bnabla \sum_{m=-\infty}
   ^\infty \int_0^\infty dr_1 r_1 K_m(r,r_1) \sigma_m(t,r_1)\exp(im\phi), 
  \eqno(\new)$$
where
$$ K_m(r,r_1) = 4\pi \sum_{\ell=|m|}^\infty {\left| Y_{\ell m}(\pi/2,
   \phi)\right|^2\over 2\ell+1} \left( {r_<^\ell\over r_>^{\ell+1}}\right)
   \approx {2\over\pi} \sum_{\ell=|m|,2}^\infty {1\over [ \ell(\ell+1.25) 
   -m^2]^{1/2}} \left[ {r_<^\ell\over r_>^{\ell+1}}\right], \eqno(\new)$$
$r_<=min(r, r_1)$ and $r_>=max(r,r_1)$. 
It can be shown that the width of $K_m$ is $r/m$, and that for $m>1$
$$ \int dr_1\, r_1 m K_m(r,r_1) \approx r. \eqno(\new)$$

If clumps were to move on circular orbits with angular speed
$\Omega(r)$, with no change to their density or shape, then the time
dependence of $\sigma_m(t,r)$ would be given by $\exp\big[im\Omega(r)t\big]$. 
We explicitly factor out this time dependence and using equation (\ref3) 
find the torque, $T$, on a cloud of mass $m_b$ to be

$$ T = 2\pi m_b G \sum_{m=-\infty}^\infty m \int_0^\infty dr_1\, r_1 K_m(r,r_1)
   \sigma_m(t,r_1)\exp\bigl[im\phi+im\Omega(r_1)t\bigr]. \eqno(\new)$$

Note that ($\phi+\Omega t$) does 
not change as we follow a blob on a circular orbit, however, the 
mutual gravitational force of clumps perturbs the orbit, and therefore 
the torque on clumps fluctuates with time causing the amplitude of their
epicyclic oscillation to undergo a random walk. We estimate the 
magnitude of the torque using equation (\last).

The torque on a cloud, in a differentially rotating disk, is dominated 
by nearby clouds. To see this, consider two rings of material at radii 
$r_1$ and $r_2$, separated by $d_{12}=|r_1-r_2|$. The torque on a cloud in
one ring due to clumps in the other ring comes from multipole moments 
$\sigma_m$ with $m\lta (r_1+r_2)/2d_{12}$. The correlation length for 
$\sigma_m$ is approximately equal to the typical size of a clump, $l_b$.
Thus the contribution from low order multipoles to the torque is small 
because within the width of the function $K_m$, $r_1/m$, we have a large
number of uncorrelated rings which give positive and negative torques
that tend to cancel each other out.\footnote{$^1$}{We are 
considering the case where there is no coherent large scale dipole or 
other low order anisotropy in the disk density distribution.} Furthermore, 
the correlation time for torque in a shearing disk falls off with distance 
as $\Omega^{-1}(d_b/d_{12}) \bigl|d\ln\Omega/d\ln r\bigr|^{-1}$ (where 
$d_b$ is the average separation between clouds and $\Omega(r)$ is the 
angular velocity).  So clouds lying at radial separation much greater 
than $d_b$ are very inefficient in transferring angular momentum, and 
most of the contribution to $T$ comes from nearby clouds.

\medskip
\ni \S2.2 Estimate of velocity dispersion and accretion rate
\smallskip

The number of clouds in a ring of radius $r$ and thickness $l_b$ 
is $\sim (r l_b/d_b^2)$, and therefore $\sigma_m\sim \sigma_b (l_b^3/ r 
d_b^2)^{1/2}$, where $\sigma_b$ is the mean surface mass density of 
clumps and $l_b$ is their mean size. It thus follows from equation
(\last) that the torque on a cloud is

$$ |T| \sim G m_b \sigma_b m_{max}\left({l_b^{2}\over d_b}\right)
\sim G m_b r \bar\sigma(r),   \eqno(\new) $$

\ni where $\bar\sigma(r)=\sigma_b (l_b/d_b)^2$ is the mean surface 
mass density of gas at $r$, and $m_{max}\sim r/d_b$. The correlation 
time for the torque $t_{cor}\sim \Omega^{-1}(d_b/d_t)^2$; where $d_t\sim
r (m_b/M_t)^{1/3}$ is the tidal radius of the cloud, and $M_t$ is the 
total axisymmetrically distributed mass contained inside the radius $r$.
Thus the change to the angular momentum of a clump in time $t_{cor}$ is 

$$\delta L \sim {G m_b^{4/3} M_t^{2/3} \over r\Omega}, \eqno(\new)$$
and the corresponding changes to the clump's radial excursion amplitude
and the radial velocity are given by

$$ \delta r \sim {G m_b^{1/3} M_t^{2/3} \over r^2\Omega^2}\sim d_b^{2/3} 
   r^{1/3} \left({M_r\over\pi M_t}\right)^{1/3} \sim d_t, \eqno(\new)$$
and
$$ \delta v_{r}\sim (r\Omega) \left[{M_{r}\over \pi M_{t}}\right]^{1/3}
   \left[{d_b\over r}\right]^{2/3}, \eqno(\new)$$
where $M_r\equiv \pi r^{2}\bar\sigma$ is the total cloud mass inside of 
radius $r$. We expect clouds to undergo collision on time scale of $t_{cor}$
when $l_b$ is not much smaller than $d_t$, and in this case the cloud 
velocity dispersion is equal to $\delta v_r$ given above.
The expression for $\delta v_r$, in the above equation, is same as that 
given by Gammie et al. (1991) for the velocity dispersion of molecular 
clouds in the Galaxy. 

Equations (\ref2) \& (\ref1) can also be obtained by considering the
gravitational interaction between clumps and the resulting epicyclic
oscillation. The change to the displacement amplitude of the epicyclic
oscillation of a clump as a result of gravitational interaction with
another clump, with impact parameter of $d\gta d_t$, can be shown to
be $\sim m_b r^3/(M_t d^2)$. Here the impact parameter is the distance of
closest approach for the unperturbed orbits of the two clumps. Gravitational
encounters with $d\lta d_t$, are almost adiabatic and there is little
change to the epicyclic energy of clumps (two particles moving on circular
orbits merely interchange their trajectories in such an encounter).
Thus the dominant gravitational interactions for 
exciting epicyclic oscillation are those with impact parameter of $\sim d_t$,
and the change to the epicyclic amplitude in such an encounter is $\delta
r \sim m_b r^3/(M_t d_t^2)\sim d_t$, which is same as the expression in
(\ref2). These encounters occur on a time interval of $\Omega^{-1}
(d_b/d_t)^2\sim t_{cor}$, and so long as the cloud size ($l_b$) is not 
much smaller than $d_b$ the collision time is also of the same order i.e. 
clouds undergo physical collision on the average after having undergone
one strong gravitational encounter, and so the velocity dispersion is 
$\sim \Omega\delta r$, which is same as equation (\last).

We have neglected the drag force of the interclump medium in our calculation
which is justified because the turbulent ram force is small compared to the 
gravitational force so long as the density of the inter-clump medium is 
less than $\bar\sigma$. Or in other words the drag force is small when
the density contrast between clump and the interclump medium is a factor
of two or greater.

The cloud collisions are highly dissipative. We assume that the kinetic 
energy of relative motion of colliding clouds is dissipated and their orbit 
is circularized. Moreover, colliding clumps coalesce 
so long as their size is not so large that rotation prevents the merger. 
Clouds are likely to fragment when their size exceed the maximum length
for gravitational instability, $l_{max}\sim(rM_r)/M_t$, preventing them
from continuing to grow in size when they undergo collision. In any case 
there is empirical evidence that gas in our galaxy, and in the 
disk of mega-maser galaxies, is clumpy (see \S3). The case considered below
is when collision between clumps are frequent so that the amplitudes of 
their epicyclic excursion do not grow to be too large.

\medskip
\noindent{\S2.2.1 Accretion rate}
\medskip

The characteristic time for a clump to fall to the center can be estimated 
using equation (\ref2) and is given by

$$ t_r \approx \Omega^{-1} \left({d_b\over r}\right)^2 \left({M_t\over m_b}
   \right)^{4/3}, \eqno(\new)$$
\ni and the average mass accretion rate is

$$ \dot M \approx \Omega(r) M_r \left({M_r\over \pi M_{t}}\right)^{4/3} 
   \,\left({d_b\over r}\right)^{2/3}. \eqno(\new)$$
As pointed out earlier, this result applies so long as the density contrast 
between the clump and the inter-clump medium is about a factor of two 
or larger, and $l_b$ is of order $d_b$. 

For $l_b\ll d_b$, clouds undergo several strong gravitational encounters 
with other clouds with impact parameter $\sim d_t$, before undergoing
a physical collision, and in this time duration, $t_{coll}\sim\Omega^{-1}
d_b^2/(d_t^{3/2} l_b^{1/2})$, their epicyclic velocity amplitude,
or the velocity dispersion, becomes 

$$\delta v_r \sim \Omega d_t \left({d_t\over l_b}\right)^{1/2}\sim (r\Omega) 
  {d_b\over (r l_b)^{1/2}} \left({M_r\over\pi M_t}\right)^{1/2}, \eqno(\new)$$
and the resulting mass accretion rate is 
$$\dot M\sim (\Omega M_r) {d_b\over (r l_b)^{1/2}}\left({M_r\over\pi M_t}
   \right)^{3/2}. \eqno(\new)$$
Since the collision time in this case is much greater than $\Omega^{-1}$, 
the effective viscosity is suppressed by a factor of $(\Omega t_{coll})^2$
(see Goldreich and Tremaine 1978, for a discussion). The physical derivation
given above automatically included this factor. Equations (\ref2) \& (\ref1)
are inapplicable in the limit $l_b\rightarrow 0$, when the collision
time becomes very long and epicyclic amplitudes saturate at a value so
that the Toore Q-parameter is about 1.

For $l_b\sim d_b\sim l_{max}$, the accretion rate in both of the cases 
considered above reduces to

$$ \dot M \approx \Omega(r) M_r \left({M_r\over \pi M_{t}}\right)^2, 
    \eqno(\new)$$
and the cloud velocity dispersion is given by
$$ v \approx r\Omega \left({M_r\over \pi M_t}\right). \eqno(\new)$$

\ni Note that the relative velocity 
of collision between clouds is smaller than the orbital speed by a factor 
of the ratio of the total mass to the molecular mass, and for $M_r\ll M_t$
collision speeds are marginally supersonic. The cooling time for gas 
depends on the density, composition and temperature, and it must be
short or comparable to the time between collisions in order for the
system to be in steady state; a specific case is discussed in \S 3.2.

We see from equation (\ref6) that for $d_b\sim l_{max}$ 
the effective kinematic viscosity is approximately 
$\l_{max}^2\Omega\sim Q^{-2} H_z^2 \Omega$, where $Q\sim l_{max}/H_z$ is 
the Toomre $Q$-parameter, $H_z\sim v/\Omega$. This effective viscosity is
same as in the ansatz suggested by Lin \& Pringle (1987) for 
self-gravitating-disk. We also note that the effective
$Q$ of the disk, for the velocity dispersion given in equation (\last), is
of order unity, and thus our solution is internally self consistent.

\bigskip
\centerline{\bf 3. Some Applications}
\medskip
In this section we apply the results of \S2 to molecular clouds within the
inner kpc of our galaxy (\S3.1), molecular disks in mega-maser 
galaxies (\S3.2), and the evolution of AGNs (\S3.3). In all these cases,
we estimate an accretion rate and compare it with previously proposed
specific model of the system.

\medskip
\centerline{\S3.1 Molecular clouds in the Galaxy}
\smallskip

As a final example we consider molecular clouds within the inner 
kilo-parsec of our own galaxy.
The amount of gas within the inner kiloparsec of our galactic center is 
$\sim 10^8$ M$_\odot$. From the molecular and HI line emissions we 
know that the gas distribution is very lumpy and much of it is contained in 
molecular clouds of radius about 10 pc. The total stellar mass within the inner 
kiloparsec is about 2x10$^{10}$ M$_\odot$. Thus using equation (\ref2) we 
estimate the accretion rate to be $\sim 10^{-5}$ M$_\odot$ yr$^{-1}$ which is
consistent with the value of $10^{-5}\alpha$/year obtained by  
Narayan et al. (1995) from a fit to the observed spectrum of Sagitarrius A*,
from radio to hard x-ray wavelengths, in their advection dominated flow model.

\medskip
\centerline{\S3.2 Mega-maser galaxies}
\smallskip

Water maser emission has been detected from about a dozen galaxies, of 
which NGC 4258 is perhaps the best studied case (Braatz, Wilson \& 
Henkel, 1996).  The maser spots in NGC 4258 fall almost along a 
straight line in the sky from which it is inferred that these sources 
lie in a thin disk the width of which is observationally estimated to 
be less than about\footnote{$^2$}{The disk thickness is expected to be 
$\sim 0.002$ pc if the molecular gas at a temperature of 10$^3$ K is 
in vertical hydrostatic equilibrium.} 0.003 pc, and the radii of the 
inner and the outer edges of the disk are 0.13pc and 0.24pc 
respectively.  From the observed acceleration of masing spots the 
central mass is inferred to be 3.6x10$^7$ M$_\odot$ (Mayoshi et al.  
1995, Greenhill et al.  1996).  The dynamical range of observation for 
NGC 4258 is about 10$^3$ and the maser amplification factor is 
estimated to be $\sim 10^4$ (cf.  Greenhill et al.  1995, Herrnstein 
et al.  1997), from which we infer that the density contrast between 
the maser spots and the mean disk is about a factor of 
three.\footnote{$^3$} {There is considerable uncertainty in estimating 
the density contrast from maser emission because the amplification 
factor for the extragalactic masers, if unsaturated, is highly 
uncertain (the density contrast would be $\sim 2$ if the gain factor 
for NGC 4258 were to be 10$^8$) and because the observed maser flux 
depends on the path length through the disk over which the line of 
sight velocity is nearly constant.  A density contrast of 2 to 3 is 
obtained by assuming that the effective path length does not vary 
rapidly across a maser spot, which is reasonable, but for which we do 
not have independent observational support.  However, we note that if 
the fluid velocity were to vary on the length scale of a spot size 
then it will necessarily lead to significant density inhomogeneity.} 
The linear drift in the observed velocities of the systemic spots in 
NGC 4258, over a period of almost 10 years, is direct evidence that 
the masing spots are discrete clumps which maintain their identity as 
they move on circular orbits (Haschick et al.  1994, Greenhill et al.  
1995, Miyoshi et al.  1995).  The density of the clouds must be at 
least 1.5x10$^{10}$ cm$^{-3}$ in order that the masing spots are not 
disrupted by the tidal field of the central mass.  This is less than 
the maximum density of 10$^{11}$ cm$^{-3}$ at which the upper energy 
state of H$_2$O is collisionally de-excited and the maser 
amplification is quenched (cf.  Reid \& Moran 1988, Greenhill et al.  
1995).  The total molecular mass of the disk, using the above 
parameters, is thus inferred to be about 2x10$^5$ M$_\odot$.  The size 
of the maser spots, $\sim 10^{-3}$ pc, is approximately the maximum 
allowed size for gravitational instability ($l_{max}$), which is 
consistent with the suggestion made in the last section that clumps 
grow in size when they undergo collision until fragmentation limits 
their size to $\sim l_{max}$.

Using equation (\last), we find that the mean random velocity of 
clumps in NGC 4258 molecular disk should be $\sim 2$ km s$^{-1}$, 
which is consistent with the scatter in the observed velocity after 
the best fitting Keplerian velocity has been subtracted from the data 
(Mayoshi et al.  1995).  Cloud collisions at this speed are barely 
supersonic (the gas temperature is expected to be between 500 and 
10$^3$ K) and molecules are not dissociated, nor do we expect much 
density enhancement in such a collision.  The time between collisions 
can be shown to be about $(d_b/l_{max})^{2/3}$ times an orbital period 
provided that $l_b\not\ll l_{max}$.  Thus we expect a cloud to undergo 
a collision once every few orbits; these collisions last for about 
$\Omega^{-1} (l_b M_t/r M_r)\sim$ a few orbits (here $l_b$ is the 
cloud size).  Colliding clouds lose energy via rotational and 
vibration transitions of several molecules of which H$_2$O and CO are 
particularly important for the NGC 4258 disk.  Using the cooling 
functions computed by Neufeld and Kaufman (1993) we find that the 
H$_2$O rotation transitions dominate the cooling at temperature of 
$\sim 10^3$K and H$_2$ number density of $\sim 10^{10}$ cm$^{-3}$, 
inspite of the fact that these lines are optically thick, and the 
resulting cooling time for colliding clouds is a few tens of years 
which is smaller than the time between collisions.  The average energy 
loss rate resulting from cloud collision in NGC 4258 disk is $\sim 
10^{38}$ erg s$^{-1}$.

We estimate the mass accretion rate for NGC 4258 using equation 
(\ref2) and find it to be 1.5x10$^{-3}$ M$_\odot$ per year.  If the 
entire gravitational potential energy release associated with this 
mass accretion rate is radiated away then the resulting bolometric 
luminosity should be about 2x10$^{43}$ erg s$^{-1}$, which is much 
greater than the observed x-ray luminosity in 2-10 kev band of 
2x10$^{40}$ erg s$^{-1}$ and the upper limit to the optical luminosity 
of 10$^{42}$ erg s$^{-1}$.  However, the observed luminosity is 
consistent with the accretion rate estimated here if most of the 
gravitational energy release of the infalling gas is advected into the 
black hole as suggested by Lasota et al.  (1996).  For a general 
discussion of the advection dominated accretion flow model, see 
Narayan and Yi (1994 \& 1995).

The H$_2$O maser emission from NGC 1068 is confined to a disk between 
radii 0.65 pc and 1.1 pc, and the mass of the central blackhole is 
estimated to be about 1.5x10$^7$ M$_\odot$ (Greenhill \& Gwinn, 1997); 
the uncertainty in these parameters, however, is larger than for the 
NGC 4258 system. If the density of the molecular gas in the disk is 
similar to that of NGC 4258, then the disk mass should be about 10$^6$ 
M$_\odot$ and the peculiar velocity of clumps, determined from 
equation (\last), comes out to be $\sim 10$ km s$^{-1}$ which is close 
to the scatter observed in the velocity of the maser knots. Equation 
(\ref2) yields an accretion rate of 0.05 M$_\odot$ yr$^{-1}$, which is 
consistent with the observed total luminosity of $\sim 4\times
10^{44}$ erg s$^{-1}$ (Pier et al. 1994) provided that the 
gravitational energy release of the gas is efficiently radiated away.

\medskip
\centerline{\S3.3 Activity in galactic centers}
\smallskip

The nuclear activity in the centers of galaxies (AGNs) is believed to 
be caused by gas accretion onto a massive central black hole.  There 
is some evidence that star formation and nuclear activity in the inner 
regions of galaxies can be triggered or enhanced by the presence of a 
close companion.  Close tidal interactions can remove sufficient 
angular momentum and enable gas to fall from the outer parts of the 
galaxy to within about 1 kpc of the center. However, tidal interactions are 
ineffective at smaller distances.  If the gas within this region is 
cold and clumpy, as for instance is the case in our own galaxy and 
some quasar hosts\footnote{$^4$}{Recent CO observation of the 
cloverleaf quasar at z=2.56 suggests that there is about 4x10$^9$ 
M$_\odot$ of molecular gas within 800 pc of the center (Barvainis 
1997).}, then, as discussed in \S2, gravitational interaction among 
clumps cause some fraction of the gas to fall inward. Since 
the accretion rate due to this process scales as $\Omega M_r (M_r/M_t)^2$
(where $\Omega$ is the angular velocity at $\sim$ kpc, and M$_r$ \&
$M_t$ are the molecular and the total mass contained within a kpc),
the accretion decreases rapidly as the gas is depleted provided that 
the standard turbulent viscosity is unimportant at $\sim $ kpc scale.  
As the fractional gas mass decreases with time (part of the gas falls 
into the blackhole and some fraction is converted into stars), the 
accretion rate falls significantly below the Eddington value and in 
the advection dominated regime where most of the gravitational energy 
release is carried with the gas inside the horizon (Narayan and Yi, 
1995) causing the luminosity of AGNs to evolve more rapidly than the 
cubic dependence of accretion rate on disk mass.  
Assuming that the star formation efficiency in the central kpc of active 
galaxies is same as the giant molecular clouds in our galaxy, we find 
the time scale for quasar luminosity evolution to be about 5x10$^8$ yrs.

The rapid evolution 
of the AGN luminosity function with redshift could perhaps be a 
consequence of the accretion dominated by the gravitational interaction 
between clumps. Since the accretion rate, due to this process, falls off as 
inverse square of the blackhole mass we expect quasars with more 
massive black holes to have a shorter activity life, provided that gas 
is converted into stars at a fixed rate and therefore more of the gas 
mass is used up in stars instead of fueling the central activity.

\bigskip
\centerline{\bf 4. Conclusion}
\medskip

We have calculated gas accretion, in a cold clumpy molecular disk, due 
to gravitational interaction among the clumps which causes fluctuating 
torque on clouds. Molecular clumps, unlike stellar systems, undergo frequent 
highly inelastic collisions which damps their epicyclic oscillation and causes 
a net inward flow of gas. We find that the accretion rate due to gravitational 
interaction among clumps is $\sim\Omega M_{mc} (M_{mc}/\pi M_{t})^2$ 
when clumps have size $\sim l_{max}$; here $l_{max}$ is the maximum 
length for gravitational instability in a shearing disk, $M_{mc}$ is the 
gas mass of the disk and $M_t$ is the sum of the central and disk mass. 

We have applied this result to water mega-maser galaxies, 
and find that the accretion rate for NGC 4258 should be about 
1.5x10$^{-3}$ M$_\odot$ yr$^{-1}$ which supports the advection 
dominated accretion flow model for this system suggested by Lasota et 
al. (1996) i.e.  most of the gravitational energy release of the 
accreting gas is advected with the flow instead of being radiated away. 
Maoz and McKee (1997) have arrived at a similar accretion rate in their 
model of NGC 4258 by assuming spiral shock in the circumnuclear disk.

The total mass in molecular clouds within the central kpc of our 
galaxy is 10$^8$ M$_\odot$. We find that the gravitational interaction 
among these molecular clouds gives rise to an accretion rate of 
$\sim 10^{-5}$ M$_\odot$ yr$^{-1}$, which is consistent with the 
value obtained by Narayan et al. (1995) from a fit to the spectrum 
of Sagitarrius A$^*$ in their advection dominated model.

If the accretion rate in AGNs were to be set by the gravitational 
viscosity discussed here (at a distance of about 1 kpc from the 
center), then their luminosity would undergo vary rapid evolution, as
observed, since the accretion rate is proportional to $M_{mc}^3$.

We have assumed in this paper that the disk is not subject to 
large scale dipole or other low order harmonic instability that can 
otherwise efficiently redistribute angular momentum and lead to 
accretion. Moreover, we have assumed that cloud size does not grow 
monotonically when they undergo collision; clouds fragment when 
their size exceeds $l_{max}$. Thus a combination of merger and 
fragmentation could maintain a steady state distribution of molecular 
clump size. We have also neglected pressure forces and other hydrodynamical
effects of cloud interaction. A fully self consistent calculation, without 
these assumptions, is desirable and perhaps will have to be attempted 
using a numerical hydrodynamical simulation of molecular disks. 

\medskip I am very grateful to John Bahcall, Ramesh Narayan and Charles 
Gammie for detailed comments which led to significant improvement in the 
presentation as well as the scientific content of the paper. I thank 
Charles Gammie for pointing out several references including the paper 
of Shlosman and Begelman.

\vfill\eject

\centerline{\bf REFERENCES}
\bigskip

\refind Barvainis, R., 1997, astro-ph/9701229, in Quasar Hosts, ed. D. 
Clements \& I. Perez-Fournon (Springer-Verlag)

\refind Braatz, J.A., Wilson, A.S., \& Henkel, C., 1996, ApJS 106, 51

\refind Gammie, C., Ostriker, J.P., and Jog, C., 1991, ApJ 378, 565

\ni Goldreich, P., and Tremaine, S., 1978, ICARUS 34, 227

\ni Greenhill, L.J., \& Gwinn, C.R., 1997

\refind Greenhill, L.J., Gwinn, C.R., Antonucci, R., and Barvainis, R., 1996,
   ApJ 472, L21

\refind Greenhill, L.J., Jiang, D., Moran, J.M., Reif, M.J., Lo, K.Y., and
Claussen, M.J., 1995, ApJ 440, 619

\refind Greenhill, L.J., Moran, J.M., and Herrnstein, J.R., 1997, to appear 
in ApJ (Letters)

\refind Haschick, A.D., Baan, W.A., and Peng, E.W., 1994, ApJ 437, L35

\refind Herrnstein, J.R., Moran, J.M., Greenhill, L.J., Diamond, P.J., 
Miyoshi, M., Nakai, N., and Inoue, M., 1997, ApJ 475, L17

\refind Ho, L.C., Filippenko, A.V., and Sargent, W.L.W., 1997, to appear in ApJ

\refind Lasota, J.-P., Abramowicz, M.A., Chen, X., Krolik, J., Narayan, R.,
\& Yi, I., 1996, ApJ 462, L142

\ni Lin, D.N.C., and Pringle, J.E., 1987, MNRAS 225, 607

\ni Maoz, E., and McKee, C.F., 1997, submitted to ApJ, (astro-ph/9704050)

\refind Miyoshi, M., Moran, J.M., Herrnstein, J.R., Greenhill, L.J., Nakai, N.,
Diamond, P., \& Makoto, I., 1995, Nature 373, 127

\refind Neufeld, D.A., and Kaufman, M.J., 1993, ApJ 418, 263

\refind Narayan, R., and Yi, I. 1994, ApJ 428, L13

\refind Narayan, R., and Yi, I. 1995, ApJ 444, 231

\refind Narayan, R., Yi, I. and Mahadevan, R., 1995, Nature 374, 623

\refind Paczynski, B., 1978, Acta Astr. 28, 91

\refind Pier, E.A., Antonucci, R., Hurt, T., Kriss, G., and Krolik, J., 1994,
ApJ 428, 124

\refind Reid, M.J., \& Moran, J.M., 1988, in Galactic and Extragalactic Radio
Astronomy, eds. G.L. Veschuer \& K.I. Kellerman

\refind Shlosman, I., and Begelman, M., 1987, Nature 329, 810

\refind Shlosman, I., Begelman, M., and Frank, J., 1990, Nature 345, 679

\ni Watson, W.D., and Wallin, B.K., 1994, ApJ 432, L35

\bye